\documentclass[cup6b]{cupbook}

\usepackage{graphicx,natbib}
\usepackage{multirow}
\bibpunct[, ]{(}{)}{,}{a}{}{,}
\def\ale{\mathrel{\hbox{\rlap{\hbox{\lower4pt\hbox{$\sim$}}}\hbox{$<$}}}}
\def\age{\mathrel{\hbox{\rlap{\hbox{\lower4pt\hbox{$\sim$}}}\hbox{$>$}}}}

\title[]
      {The GRB--SN connection}
\author[J. Hjorth \& J.S. Bloom]{Jens Hjorth \\
        Dark Cosmology Centre, Niels Bohr Institute, University of Copenhagen, \\
       Juliane Maries Vej 30, DK-2100 Copenhagen \O, Denmark
        \and 
        Joshua S.\ Bloom\\
        Astronomy Department, University of California at Berkeley, \\
	601 Campbell Hall, Berkeley, CA 94720, USA}

\date{7 April 2011}

\begin{document}

\pagenumbering{roman}
\pagenumbering{arabic}

\setcounter{chapter}{8}

\chapter{The GRB-Supernova Connection}

\author{Jens Hjorth \\
        Dark Cosmology Centre, Niels Bohr Institute, University of Copenhagen, Denmark
       \and 
       Joshua S. Bloom\\
       Astronomy Department, University of California at Berkeley, Berkeley, CA, USA}

\noindent 

\section{Introduction}

The discovery and localisation of the first afterglows of GRBs rapidly led to the establishment of the long-sought distance scale for the sources (see Chapter~4), which began an earnest observational hunt for the progenitors. A preponderance of evidence linked long-duration, soft-spectrum GRBs with the death of massive stars. The observations of the GRB--supernova (SN) connection, the main subject of this chapter, present the most direct evidence of this physical link.

Well before the Afterglow Era, \citet{pac86} noted that ``cosmological'' distances of GRBs would imply that the energy release in $\gamma$ rays would be comparable to the energy release in a typical SN explosion. Seen as more than just a coincidence, this energetics connection between GRBs and the death of massive stars was fleshed out\footnotemark\footnotetext{The possible connection between GRBs and SNe was actually first studied observationally in the original GRB discovery paper of \citet{kso73}, following the suggestion that $\gamma$ rays could be produced in SN shock breakout \citep{col68}.} into what is now referred to as the {\it collapsar model} \citep{1993ApJ...405..273W,1996AIPC..384..709W,1999ApJ...524..262M}. Briefly, the collapsar involves the core-collapse explosion of a stripped-envelope massive star. Matter flows towards a newly formed black hole or rapidly spinning, highly magnetized neutron star (``magnetar''; e.g., \citealt{2009MNRAS.396.2038B}). Powerful jets plow through the collapsing star along the spin-axis, eventually obtain relativistic speeds, and produce GRBs. Enough $^{56}$Ni is produced near the central compact source to power a supernova explosion of the star. 
The original  ``failed Ib'' model posited that little $^{56}$Ni would be produced during core-collapse of a massive star that produces a GRB, and thus no traditional SN would be visible. 
Chapter 10 is an in-depth review of core-collapse progenitor models, including collapsar and millisecond-magnetar-driven models.

After the first few afterglow localizations of \citep[long-duration;][]{1993ApJ...413L.101K} GRBs, a qualitative connection of the events with star-formation regions and star-forming galaxies began to emerge (e.g., \citealt{1998ApJ...494L..45P}). The close proximity of GRBs to star formation, surprising to many, was not a natural expectation of degenerate merger models (e.g., \citealt{fwh99}; \citealt{bsp99}). Instead, this evidence directly implicated models where the progenitor does not move far from its birthsite and produces a GRB on timescales smaller than the typical duration of star-formation episodes. The discovery of the energetic core-collapse supernova 1998bw associated with the very underluminous GRB\,980425 (see section \ref{sec:grb-sn:1998bw}) provided the first concrete evidence for a GRB connection with a massive star death (though the relationship between GRB\,980425, at the low redshift of $z=0.0085$, and ``cosmological'' GRBs would be debated for years). As statistical statements about the physical connection of cosmological GRBs with on-going star formation amassed \citep{bkd02,ldm+03,2004A&A...425..913C,2006Natur.441..463F}, individual events began to exhibit credible {\it photometric} evidence for a SN explosion contemporaneous with the GRB \citep{1999Natur.401..453B,2000ApJ...536..185G,2002ApJ...572L..45B,2003ApJ...582..924G,2004ApJ...609..952Z}. The definitive evidence for the GRB-SN connection was finally established by a few events which, through {\it spectroscopic} identification of SN features well after the GRB event (see section \ref{sec:grb-sn:spec}), clinched the physical association. An early review of the GRB-SN connection was given in \citet{JanvanParadijs10221999} and more recent dedicated reviews were given in \citet{2006AIPC..836..380S}, \citet{2006ARA&A..44..507W} and \citet{2007RMxAC..30..104D}.
 
We begin this chapter by reviewing the strongest current observational 
evidence for the GRB-SN connection; Chapter 10 provides an overview of how 
these observations tie, theoretically, some GRBs to the death of massive stars. 
We summarize 30 GRB-SN associations and focus on five ironclad GRB-SN 
associations (GRB\,980425/SN\,1998bw, GRB\,030329/SN 2003dh, 
GRB\,031203/SN\,2003lw, GRB\,060218/SN\,2006aj, GRB\,100316D/SN\,2010bh), 
discovered by four different satellites, which, in concert, constitute 
irrefutable spectroscopic evidence for the association of GRBs and SNe. 
We highlight the subsequent insight into the progenitors enabled by detailed 
observations. 
We next present some of the supporting evidence for the GRB-SN connection and
address the SN association (or lack thereof) with several sub-classes 
of GRBs, finding that the X-ray Flash (XRF) population is likely associated 
with massive stellar death whereas short-duration events likely arise from an 
older population not readily capable of producing a SN concurrent with a GRB. 
Interestingly, a minority population of seemingly long-duration, soft-spectrum 
GRBs show no evidence for SN-like activity; this may be a natural consequence 
of the range of $^{56}$Ni production expected in stellar deaths. We conclude 
the chapter by providing an outlook for the next decade of GRB-SN research.

\section{Spectroscopic evidence for GRBs and SNe}
\label{sec:grb-sn:spec}

\subsection{GRB\,980425/SN\,1998bw}
\label{sec:grb-sn:1998bw}

\begin{figure}
    \centering
    \includegraphics[scale=0.4,angle=90]{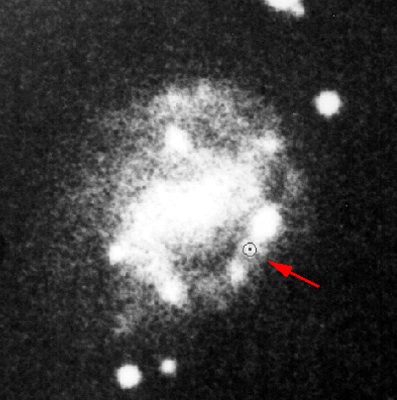}
    \includegraphics[scale=0.4,angle=90]{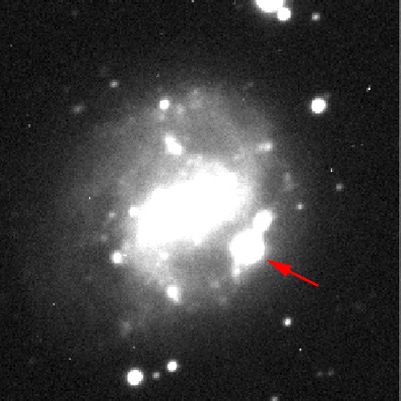}
    \includegraphics[scale=0.252]{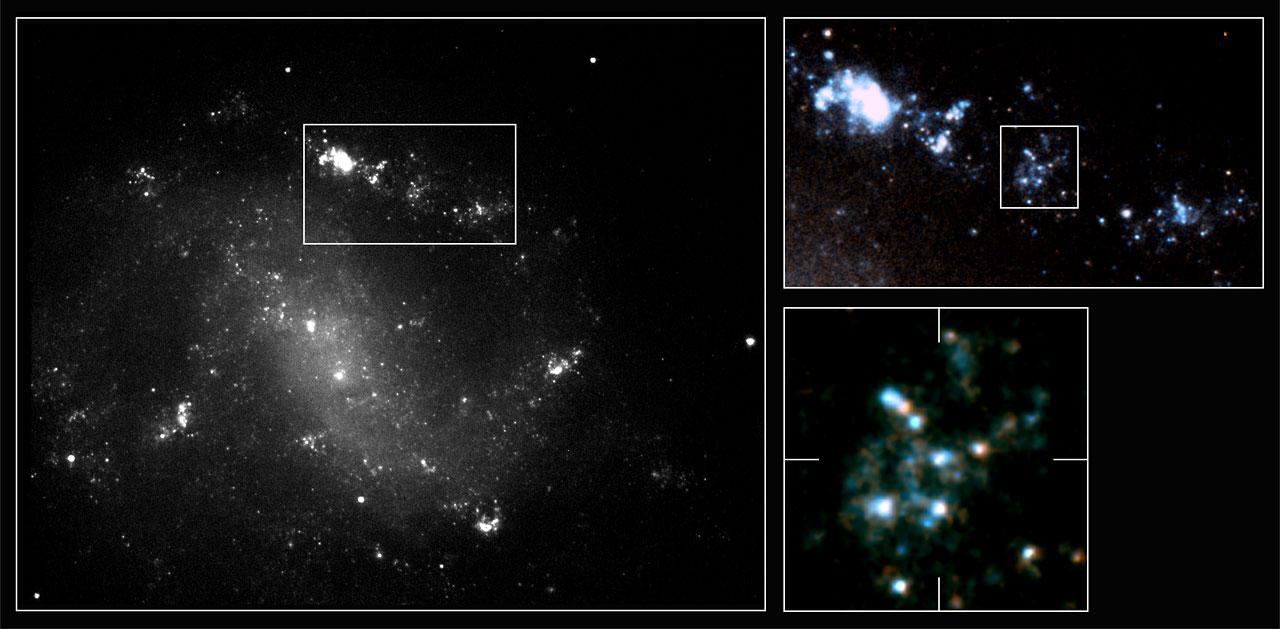}
    \caption{Discovery of SN\,1998bw associated with GRB\,980425.
    The upper panels show the images of the host galaxy of
    GRB\,980425, before (left) and shortly after (right) the
    occurrence of SN\,1998bw \citep{1998Natur.395..670G}.
    The bottom panel shows a late {\it HST\,} image of the host
    galaxy and SN\,1998w. The 3-step zoom-in shows SN\,1998bw 778 days 
    after the explosion embedded in a large star-forming region of a spiral arm \citep{2000ApJ...542L..89F}.
    }
    \label{f:1998bw_discovery}
  \end{figure}

GRB\,980425 was discovered early in the afterglow era; at the time it exploded 
only six GRBs had been localized by {\it BeppoSAX} and only two had measured 
redshifts, 
namely GRB\,970508 at $z=0.83$ \citep{1997Natur.387..878M} and GRB\,971214 at 
$z=3.42$ \citep{1998ApJ...509L...5O}. In the error circle of GRB\,980425, 
two X-ray sources were found, though the precise characterisation of their respective variability was uncertain.  Therefore, the identification of the true X-ray counterpart to GRB\,980425 was controversial \citep{2000ApJ...536..778P,2004ApJ...608..872K}; as a result, the discovery of a supernova, SN\,1998bw (Fig.~\ref{f:1998bw_discovery}), coincident with one of the X-ray sources was not immediately taken as unequivocal evidence for a direct link to GRB\,980425.  

\begin{figure}
    \centering
    \includegraphics[scale=0.45]{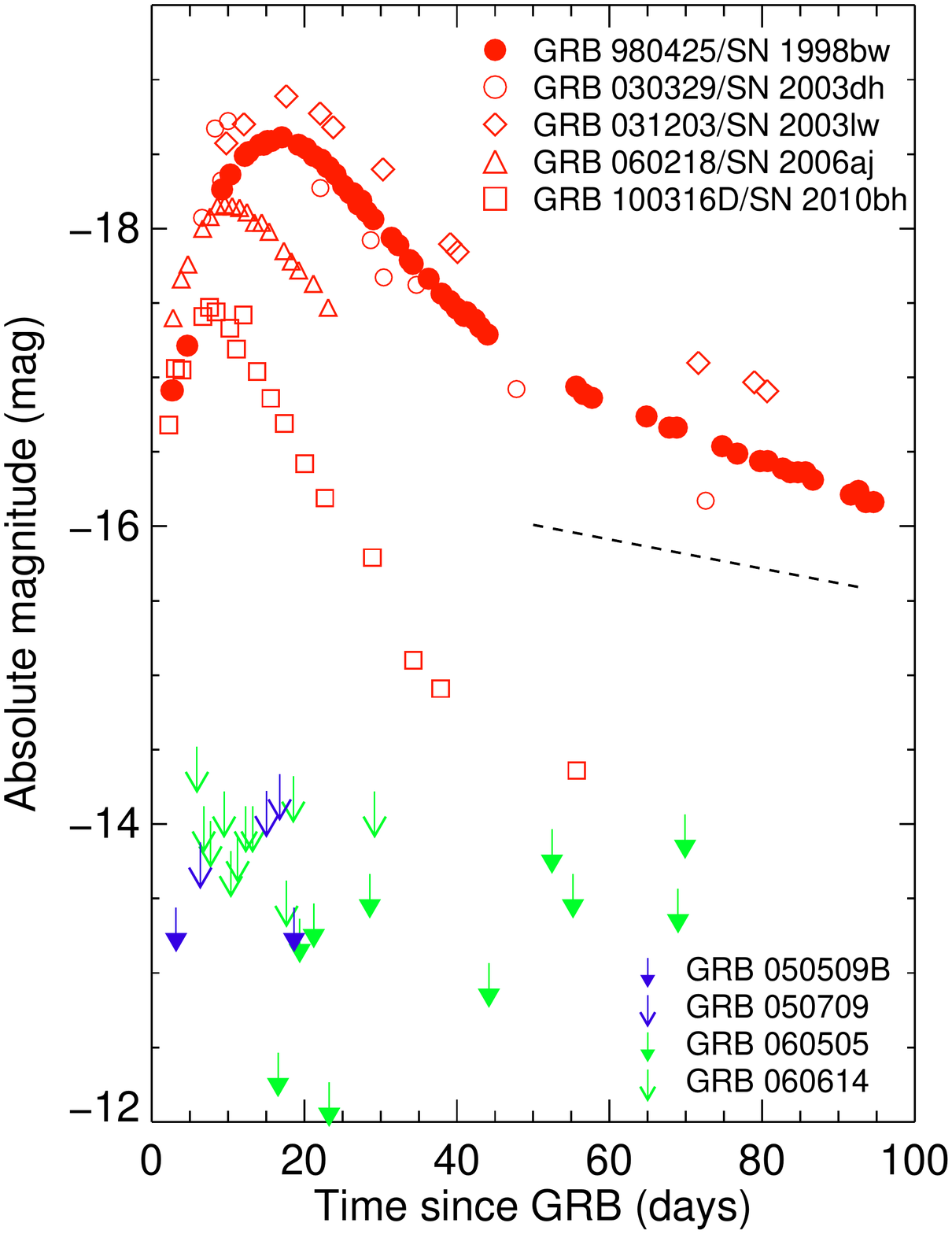}
    \caption{Light curves of spectroscopic GRB-SNe contrasted with upper 
    limits on SNe in SN-less GRBs. The red data points are the
    light curves of the GRB-SNe 
    1998bw \citep{1998Natur.395..670G},
    2003dh \citep{2003Natur.423..847H},
    2003lw \citep{2004ApJ...609L...5M},
    2006ap \citep{2006Natur.442.1011P} (bolometric magnitudes adapted from
    \citealt{2006Natur.442.1011P} who used a cosmology with
    $H_0=73$ km s$^{-1}$ Mpc$^{-1}$, $\Omega_\Lambda=0.72$, $\Omega_m=0.28$),
    and 2010bh \citep{2011AN....332..262B}. 
    The upper limits are from the short GRBs
    050509B \citep{2005ApJ...630L.117H}
    and 050709 \citep{2005Natur.437..859H}
    (blue arrows) 
    and two SN-less long GRBs (green arrows) 
    \citep{2006Natur.444.1047F}.
    Approximate bolometric magnitudes are based on R and V band upper limits 
    offset relative to the
    corresponding SN 1998bw V or R band light curves. Time is in the restframe.
    The $^{56}$Co decay slope is shown for reference (dashed curve).
    }
    \label{f:lightcurves}
  \end{figure}

\begin{figure}
    \centering
    \includegraphics[scale=0.55]{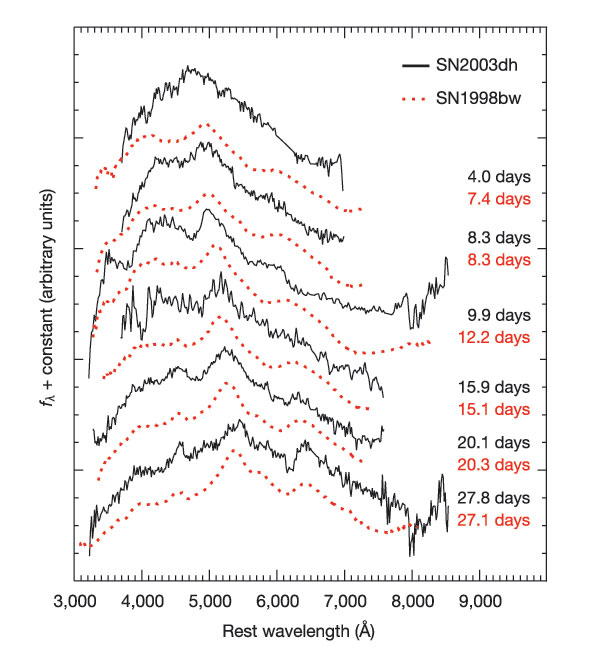}
    \caption{Spectral evolution of GRB-SNe 1998bw 
    \citep{2001ApJ...555..900P}
    and 2003dh \citep{2003Natur.423..847H}.
    Solid lines indicate spectra of SN 2003dh obtained by subtracting a model
    for the afterglow and host galaxy contributions from the spectra.
    Dotted red lines indicate spectra of SN 1998bw taken at similar epochs. 
    Times after the GRB are indicated in the rest frame. From \citet{2003Natur.423..847H}.
    }
    \label{f:2spectra}
  \end{figure}

SN 1998bw was a spectacular event. It was a bright ($M_B=-18.7$\,mag 
at peak), broad-lined Type Ic SN \citep{1998Natur.395..670G} suggesting a significant amount of mass with very fast (upwards of $30,000$ km s$^{-1}$) photospheric expansion (\citealt{2006ARA&A..44..507W} advocate for the designation as Ic-BL, for broad-lined SN without He, H or Si in the spectrum). The light curve is shown in Fig.~\ref{f:lightcurves} and its spectral evolution 
\citep{2001ApJ...555..900P} is shown in Fig.~\ref{f:2spectra}. The late-time light curve, extending to 500 days after the GRB, exhibited a decline consistent with cobalt decay to iron \citep{1999PASP..111..964M,2000ApJ...537L.127S}. At the time, 
SN\,1998bw was also the brightest radio SN known, 
indicating, as a means to explain the very high apparent brightness 
temperature, that the SN was accompanied by a shock wave moving at mildly 
relativistic speeds (\citealt{1998Natur.395..663K}; but see also 
\citealt{1999ApJ...515..721W}). 
\citet{1998Natur.395..672I} suggested that these observations can be reproduced 
by an extremely energetic explosion of a massive star composed mainly of carbon 
and oxygen (having lost its hydrogen and helium envelopes). Based upon an independent modelling effort, \citet{1999ApJ...516..788W} concurred with the carbon and oxygen core hypothesis and also argued that SN\,1998bw was an asymmetric explosion. \citet{1998Natur.395..672I} and others at the time used the term 
``hypernova'' \citep{1998ApJ...494L..45P} to describe such a very energetic SN, modelled to have released roughly 10 times more energy than in a typical (10$^{51}$ erg) SN. We caution here that the term ``hypernova'' is a theory-laden classification pertaining to energetics; it is entirely possible to have a core-collapse SN with large expansion velocity ($\age 20,000$ km s$^{-1}$) yet typical ($10^{51}$ erg) energy coupled to the ejecta.

No traditional optical afterglow (as seen in most
other GRBs; see Chapters 4--6) was detected. Moreover, the comparatively low energy output of GRB\,980425 \citep[see e.g.,][]{2007ApJ...654..385K} and its low redshift were considered as pointing to a different class of GRB
\citep{1998Natur.395..663K,1998ApJ...506L.105B}, not necessarily of the same progenitor origin as the truly cosmological GRBs (loosely defined as having a significant redshift, a high energy output in $\gamma$ rays, $E_\gamma \sim 10^{52}$ erg, and an (optical) afterglow decaying as a power law) that had been detected so far.

Doubts therefore remained about the GRB-SN connection, arising from an {\it a posteriori} statistical argument about the association in time and place with two X-ray sources in the $\gamma$-ray error circle, the lack of an optical afterglow, and the  low implied 
energy output; even if the physical connection was believed, GRB\,980425 was clearly set apart from the typical cosmological GRBs emitting orders of magnitude more $\gamma$-ray energy (Table~\ref{t:five}).

\subsection{GRB\,030329/SN\,2003dh}

Though supporting evidence for a GRB-SN connection grew in the interim (see section \ref{sec:support}), almost 5 years after GRB\,980425/SN\,1998bw, GRB\,030329 eliminated any doubts as to the deep connection of the two phenomena. GRB\,030329 was a bright burst detected by the {\it HETE-2} satellite \citep{2004ApJ...617.1251V,2004ApJ...606..381L}.
At an inferred redshift of $z=0.1685$ \citep{2003GCN..2020....1G}, it was 
``truly'' cosmological and had a total isotropic
energy release $10^{4}$ times that of GRB\,980425. Moreover, it was followed by a 
bright optical afterglow \citep{2003Natur.423..844P}, helping solidify this event as part of the cosmological GRB class. Several days after the 
burst, the optical spectrum started to change from a featureless power-law 
spectrum, characteristic of GRB afterglows, to include more and more SN 
features \citep{Matheson03_GCN2107,2003ApJ...591L..17S,2003Natur.423..847H,2003ApJ...593L..19K,2003ApJ...599..394M}.
By subtracting the afterglow contribution,
the SN spectrum could be isolated. It 
was shown to closely follow that of SN\,1998w, thus conclusively showing that 
the GRB afterglow and SN were spatially coincident and that GRB\,030329 and 
SN\,2003dh were co-eval to within a few days \citep{2003Natur.423..847H} (Fig.~\ref{f:2spectra}). 
Lingering possibilities\footnote{Favored by the now questionable identification of spectral lines in X-ray afterglow spectra. See also Chapters 4 and 5.} of a 
SN preceding the GRB by years to weeks 
\citep[the so-called ``supranova'' model; ][]{1998ApJ...507L..45V}
were all but ruled out.

The SN light curve was almost completely masked by the bright afterglow
\citep{2004ApJ...606..381L}.
Only by subtraction of the afterglow was it obvious that SN\,2003dh peaked at 
about the brightness of SN\,1998bw but evolved faster
\citep{2003Natur.423..847H,2003ApJ...599..394M}
(Fig.~\ref{f:lightcurves}).

In retrospect, the GRB\,030329/SN\,2003dh connection also eliminated any doubts
about the association between GRB\,980425 and SN\,1998bw.
GRB 030329 remains the only GRB with both a clear optical 
afterglow and a convincing spectroscopically confirmed SN. 

\subsection{GRB\,031203/SN\,2003lw}

GRB\,031203 was localized by {\it Integral} \citep{2004Natur.430..646S} and 
the afterglow was subsequently accurately 
localized by {\it Chandra} \citep{2004ApJ...609L..59G}, {\it XMM-Newton}
\citep{2004ApJ...603L...5V}, and the VLA \citep{2004ApJ...609L..59G} to a galaxy 
at $z = 0.1055$ \citep{2004ApJ...611..200P}. No optical afterglow was detected,  
but through photometric monitoring of the galaxy a SN light curve bump 
was detected 
\citep{2004A&A...419L..21T,2004ApJ...608L..93C,2004ApJ...609L..59G}.

\begin{figure}
    \centering
    \includegraphics[scale=0.59]{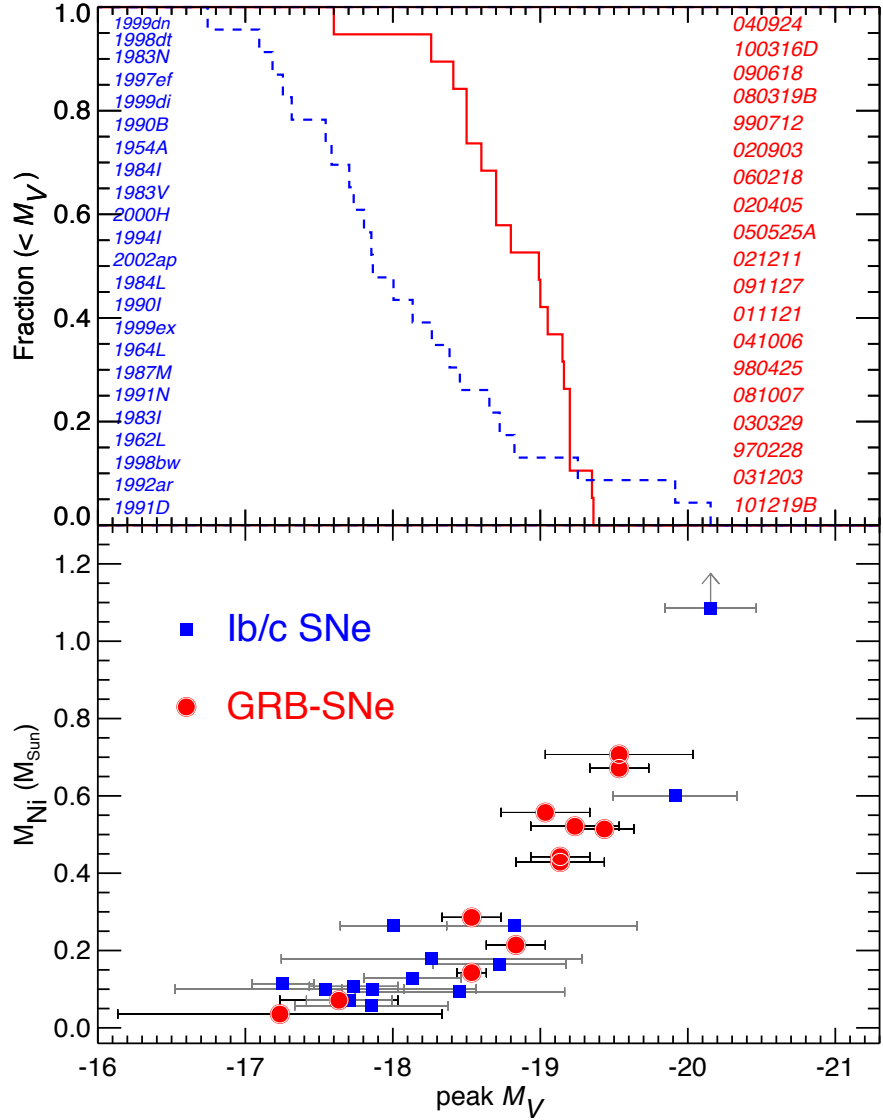}
   \caption[]{\small ({\it Top}) Comparison of the peak brightness of SNe related to GRBs/XRFs (solid cumulative histogram) with local Ibc SNe with well-measured peak brightnesses (dashed histogram); updated from \citet{2006ARA&A..44..507W}.  Historical Type Ib/c SNe are listed on the left. Only those events with a ``C'' grade or better from Table \ref{t:all} are shown. To mitigate the numerous biases in finding SNe associated with GRB/XRFs, \citet{2006ARA&A..44..507W} considered as upper limits the non-detections from the literature and bump claims with less firm SN detections, finding a statistical consistency of peak brightness distributions of the two populations. ({\it Bottom}) Comparison of implied $^{56}$Ni mass and {\it modelled} peak brightnesses between GRB-SNe and Ibc SNe (adopted for $h=0.71$ cosmology from \citealt{2006AJ....131.2233R,2009AJ....137..347R}). Note: the Richardson model peak brightnesses can differ from the literature by $\approx$0.5 mag.}
    \label{f:skdiagram}
  \end{figure}

Spectra of the SN obtained by \citet{2004ApJ...609L...5M} revealed 
broad-line features similar to those seen in SN\,1998bw and SN\,2003dh.  
While the peak brightnesses of SN\,1998bw and SN\,2003dh were similar, SN\,2003lw
was brighter by about 0.3--0.5 mag at peak; 
the uncertainty in the intrinsic brightness reflects a
large Galactic extinction towards the field, but SN\,2003lw nevertheless
provides strong evidence that GRB SNe are not standard candles (see Fig.~\ref{f:skdiagram}). 

While the large column density of Galactic dust was a nuisance --- dimming and 
reddening the SN and the afterglow --- it  allowed for a novel 
measurement: two radially expanding halos of X-ray emission centered on the GRB 
were discovered in {\it XMM-Newton} observations \citep{2004ApJ...603L...5V}.
Because of the time delay due to the longer distance traveled by the light in 
the rings this allowed a reconstruction of the prompt X-ray flux of GRB\,031203
\citep{2004ApJ...605L.101W,2006A&A...449..203T,2006ApJ...636..967W}.
It was suggested that the fluence in X rays dominates that of the 
hard-energy emission (though see \citealt{2004ApJ...611..200P}); in other words, had a hard X-ray detector observed the prompt emission of GRB\,031203, it might have been classified as an X-ray flash 
(XRF: a dominant fraction of the total prompt fluence are detected as X rays rather than $\gamma$ rays; \citealt{hzk+01}). Therefore, we take SN\,2003lw as reasonable spectroscopic evidence for an 
XRF-SN connection.

While \citet{2004ApJ...609L...5M} reported evidence for a very red, faint 
infrared afterglow, GRB\,031203 was essentially SN dominated. 
Consequently, \citet{2004A&A...419L..21T} speculated that {\it Swift} would not 
only be a GRB mission, but also a discoverer of SNe as they explode. This prediction came to fruition with GRB\,060218/SN\,2006aj. 

\subsection{GRB\,060218/SN\,2006aj}

GRB\,060218 was localized 
by the {\it Swift} satellite to a galaxy at $z=0.0335$ \citep{2006ApJ...643L..99M}. \citet{2006Natur.442.1008C} and \citet{2007ApJ...667..351W}
argued that {\it Swift} detected emission due to the shock breakout of the 
associated SN (see, however, \citealt{2007ApJ...656.1001B}
and \citealt{2007MNRAS.382L..77G}). SN\,2006aj was fainter than the other GRB SNe, providing additional evidence for a substantial dispersion in the peak 
magnitudes and rise times of GRB SNe
\citep{2006Natur.442.1011P,2006ApJ...645L..21M,2006A&A...454..503S,2006A&A...457..857F}
(Fig.~\ref{f:lightcurves}).
Building on the 030723 \citep{2004ApJ...609..962F},
020903 \citep{2005ApJ...627..877S} and 031203 hypothesis of 
an XRF-SN connection, GRB\,060218 was classified as an XRF (though the duration [$>1000$ s] of the event was very long compared to other XRFs; \citealt{2006Natur.442.1008C}).

\subsection{GRB\,100316D/SN\,2010bh}

The fifth and latest known {\it bona fide} spectroscopic
GRB-SN, GRB 100316D/ SN\,2010bh, was
also discovered by {\it Swift} \citep{2011MNRAS.411.2792S}.
At a redshift of $z = 0.0591$, it had high-energy prompt properties that 
were remarkably similar to those of GRB\,060218:
it was also an XRF of unusually long duration
(1300 s), it had a thermal component in addition to a synchrotron 
emission component with a low peak energy, a slow X-ray decay,
and similar spectral hardness evolution.

The supernova features were typical of broad-lined SNe Ic and
generally consistent with other spectroscopic GRB-SNe
\citep{2010arXiv1004.2262C}. The Si II $\lambda$6355 
expansion velocity was much higher than for SN\,2006aj, more
similar to SN\,1998bw and SN\,2003dh 
\citep{2010arXiv1004.2262C,2001ApJ...555..900P,2003Natur.423..847H}.
GRB\,100316D/SN\,2010bh was also distinct from GRB\,060218/SN\,2006aj in that
their host galaxies were
different, with the host of GRB\,100316D more closely resembling that 
of GRB\,980425 \citep{2011MNRAS.411.2792S}.

\section{Supporting evidence for the GRB-SN connection}
\label{sec:support}

The highest redshift among the secure GRB-SNe is $0.1685$ while the median 
redshift of {\it Swift} GRBs is above 2 \citep{2006A&A...447..897J}. Moreover, save 
GRB\,030329, all of 
the spectroscopically secure events released significantly less energy than the median $E_\gamma$ of known
cosmological GRBs \citep{2007ApJ...654..385K}.
It is, therefore, important to consider the evidence for GRB SNe at higher 
redshifts and for GRBs with  $E_{\gamma,\rm iso}$ in the range
$10^{50}$--$10^{54}$ erg \citep{2001ApJ...562L..55F}.
At higher redshift secure SN identification becomes difficult because the SN 
appears fainter, which leads to the difficulty of obtaining a sufficient signal-to-noise ratio in the broad SN features. The signal-to-noise problem  is aggravated by
the contamination of the host galaxy and the afterglow, which do not necessarily
get comparatively fainter with redshift \citep[see, e.g.,][]{2006ARA&A..44..507W}.

There exists, however, substantial
photometric evidence for late-time light\-curve bumps. 
The first clear-cut case for a bump was GRB\,980326 
\citep{1999A&AS..138..449C,1999Natur.401..453B},
quickly followed by reanalysis of GRB\,970228 
\citep{1999ApJ...521L.111R,2000ApJ...536..185G}
and GRB\,990712 \citep{2001ApJ...552L.121B}. Spectroscopic confirmation was 
attempted in several other systems with secure bumps, including 
GRB\,011121 \citep{2003ApJ...582..924G,2003ApJ...599.1223G,2002ApJ...572L..45B},
XRF\,020903 \citep{2005ApJ...627..877S,2006ApJ...643..284B},
GRB\,021211 \citep{2003A&A...406L..33D},
XRF\,030723 \citep{2004ApJ...609..962F,2004ApJ...612L.105T},
GRB\,050525A \citep{2006ApJ...642L.103D}, 
GRB\,081007 \citep{2008CBET.1602....1D}, and
GRB\,101219B \citep{Sparre2011}, 
providing in all cases tentative evidence for SN spectroscopic features.
Other systems with clear bumps include 
GRB\,020405 \citep{2003ApJ...589..838P}, 
GRB\,041006 \citep{2005ApJ...626L...5S}, 
GRB\,080319B \citep{2010ApJ...725..625T}, 
GRB\,090618 \citep{2010arXiv1012.1466C}, and 
GRB\,091127 \citep{2010ApJ...718L.150C};
GRB\,020410 was suggested to have been discovered by its SN light
\citep{2005ApJ...624..880L}
and a bump of more dubious origin was detected in GRB\,020305 
\citep{2005A&A...437..411G}.
The bulk of the evidence points to Type Ic SNe and only
in a few cases Type II SNe have been suggested
\citep[e.g.,][]{2003ApJ...582..924G,2005A&A...437..411G}.
Only a few have IAU designations.
In Table~\ref{t:all} we give a list of their properties. We also attempt
to grade the GRB-SN connections according to the strength of the
existing observational evidence. For an independent analysis of bursts
up to 2003, see \citet{2004ApJ...609..952Z}. 

 \begin{table}
  \caption{Evidence for GRB-SNe.}
    \begin{tabular}{l|ccccl}
     \hline \hline
GRB/XRF & SN     & $z$   &Evidence&Comments             &Refs.  \\
        & Designation &   &      &   & \\
     \hline
970228  &        &0.695  &  C     &                     &1,2\\
980326  &        &       &  D     &red bump             &3,4\\
980425  &1998bw  &0.0085 &  A     &spectroscopic SN     &5\\
990712  &        &0.433  &  C     &                     &6\\
991208  &        &0.706  &  E     &low significance     &7\\
000911  &        &1.058  &  E     &low significance     &8,9\\
011121  &2001ke  &0.362  &  B     &spectral features    &10,11,12\\
020305  &        &       &  E     &not fitted by GRB-SNe&13\\
020405  &        &0.691  &  C     &red bump             &14,15\\
020410  &        &       &  D     &discovered via bump  &16\\
020903  &        &0.251  &  B     &spectral features    &17,18\\
021211  &2002lt  &1.006  &  B     &spectral features    &19\\
030329  &2003dh  &0.1685 &  A     &spectroscopic SN     &20,21,22\\
030723  &        &       &  D     &red bump, X-ray excess &23,24\\
031203  &2003lw  &0.1055 &  A     &spectroscopic SN     &25\\
040924  &        &0.859  &  C     &                     &26,27\\
041006  &        &0.716  &  C     &                     &26,28\\
050416A &        &0.654  &  D     &poor sampling        &29\\
050525A &2005nc  &0.606  &  B     &spectral features    &30\\
050824  &        &0.828  &  E     &low significance     &31\\
060218  &2006aj  &0.0334 &  A     &spectroscopic SN     &32,33,34\\
060729  &        &0.543  &  E     &afterglow dominated  &35,36\\
070419A &        &0.971  &  D     &poor sampling        &35,37,38\\
080319B &        &0.938  &  C     &multiple color bump &35,39,40,41\\
081007  &2008hw  &0.530  &  B     &spectral features    &42,43,44\\
090618  &        &0.54   &  C     &                     &36,45\\
091127  &2009nz  &0.490  &  C     &                     &46\\
100316D &2010bh  &0.0591 &  A     &spectroscopic SN     &47,48\\
100418A &        &0.624  &  D     &                     &49\\
101219B &        &0.552  &  B     &spectral features    &50,51,52\\
     \hline \hline
    \end{tabular}
  \label{t:all}
The evidence according to the authors for a SN associated with 
a GRB is listed in column (4) according to the following scale: 
A: Strong spectroscopic evidence.
B: A clear light curve bump as well as some spectroscopic evidence resembling
a GRB-SN.
C: A clear bump consistent with other GRB-SNe at the spectroscopic redshift
of the GRB.
D: A bump, but the inferred SN properties are not fully consistent with other GRB-SNe or
the bump was not well sampled or there is no spectroscopic redshift of the GRB.
E: A bump, either of low significance or inconsistent with other GRB-SNe.
\\
{\bf References}: \\
(1) \citet{1999ApJ...521L.111R}
(2) \citet{2000ApJ...536..185G}
(3) \citet{1999Natur.401..453B}
(4) \citet{1999A&AS..138..449C}
(5) \citet{2006Natur.444.1047F}
(6) \citet{2001ApJ...552L.121B}
(7) \citet{2001A&A...370..398C}
(8) \citet{2001A&A...378..996L}
(9) \citet{2005A&A...438..841M}
(10) \citet{2002ApJ...572L..45B}
(11) \citet{2003ApJ...582..924G}
(12) \citet{2003ApJ...599.1223G}
(13) \citet{2005A&A...437..411G}
(14) \citet{2003ApJ...589..838P}
(15) \citet{2003A&A...404..465M}
(16) \citet{2005ApJ...624..880L}
(17) \citet{2005ApJ...627..877S}
(18) \citet{2006ApJ...643..284B}
(19) \citet{2003A&A...406L..33D}
(20) \citet{2003ApJ...591L..17S}
(21) \citet{2003Natur.423..847H}
(22) \citet{2003ApJ...593L..19K}
(23) \citet{2004ApJ...609..962F}
(24) \citet{2004ApJ...612L.105T}
(25) \citet{2004ApJ...609L...5M}
(26) \citet{2006ApJ...636..391S}
(27) \citet{2008A&A...481..319W}
(28) \citet{2005ApJ...626L...5S}
(29) \citet{2007ApJ...661..982S}
(30) \citet{2006ApJ...642L.103D}
(31) \citet{2007A&A...466..839S}
(32) \citet{2006Natur.442.1011P}
(33) \citet{2006ApJ...645L..21M}
(34) \citet{2006A&A...454..503S}
(35) \citet{2009ApJS..185..526F}
(36) \citet{2010arXiv1012.1466C}
(37) \citet{Cenko07_GCN6322}
(38) \citet{Hill07_GCN6486}
(39) \citet{Kann08_GCN7627}
(40) \citet{bloometal09}
(41) \citet{2010ApJ...725..625T}
(42) \citet{Berger08_GCN8335}
(43) \citet{2008CBET.1602....1D}
(44) \citet{Soderberg08_GCN8662}
(45) \citet{2009GCN..9518....1C}
(46) \citet{2010ApJ...718L.150C}
(47) \citet{2010arXiv1004.2262C}
(48) \citet{2011AN....332..262B}
(49) A. De Ugarte-Postigo (private communication, 2010)
(51) \citet{2011GCN.11578....1O}
(50) \citet{DeUgartePostigo11_GCN11579}
(52) \citet{Sparre2011}

\end{table}

Generally, the picture supports the conclusion of the previous section, namely 
that SNe are ubiquitous in GRB light curves and that there is real diversity 
among such events
\citep{2004ApJ...609..952Z,2006A&A...457..857F,2006ARA&A..44..507W,2009AJ....137..347R}.
GRB-SNe are generally consistent with being broad-lined Type Ic, with
a dispersion in both peak brightness, rise time, light curve width, and
spectral broadness.  
They appear to represent the brighter end of the Ib/Ic population 
(see Fig.~\ref{f:skdiagram}), but \citet{2006ARA&A..44..507W} showed that when the non-detections of GRB-SNe are accounted for, the two populations are statistically consistent with having been drawn from the same population. Clearly, as rare events, very bright Ic SNe are not routinely identified, nor are the fainter GRB-SNe. So as more uniform populations of GRB-SNe (and non-detections thereof) and Ib/Ic (and particularly Ic-BL) SNe are produced, it will be instructive to revisit this question of the comparative brightness and $^{56}$Ni distribution.

At radio wavebands, GRB afterglows can be 10$^4$ times brighter at peak than typical Ib and Ic SNe (see \citealt{2006AIPC..836..380S}): the reason is likely the difference between the coupling of energy to highly relativistic ($\Gamma \age 50$) ejecta (in the GRB/XRF case) versus sub-relativistic ($\beta \Gamma \ale$ 1) ejecta (in the normal Ib/Ic case). The early radio brightness of SN 1998bw can be attributed to the large coupling of energy to trans-relativistic ejecta ($\beta \Gamma \approx$ a few; \citealt{1998Natur.395..663K,2006AIPC..836..380S}).

\section{Supernova-less GRBs}
\label{sec:less}

Having presented the evidence that firmly established the GRB-SN connection 
we shall now present evidence for the existence of SN-less GRBs. These 
fall in two categories.

\subsection{Short-duration GRBs}

Short-duration, hard-spectrum GRBs (see Chapters 3 and 11) consistently do 
not exhibit SN features in their optical afterglow light curves. This was 
demonstrated already for the first two localized short GRBs, GRB\,050509B 
\citep{2006ApJ...638..354B,2005ApJ...630L.117H} 
and GRB\,050709 \citep{2005Natur.437..859H,2005Natur.437..845F}
(Fig.~\ref{f:lightcurves}).

Bright transient emission, dubbed a ``mini SN'' 
\citep{1998ApJ...507L..59L,2002MNRAS.336L...7R}, could 
be produced by radioactive 
elements that are synthesized during the rapid decompression of very dense 
and neutron rich material ejected during a NS-NS or a NS-BH merger 
\citep[see, e.g.,][]{1999A&A...341..499R}. The emission is expected to 
peak around 
the optical-UV range within a day or so with a semi-thermal spectrum
\citep{1998ApJ...507L..59L}.  
Since modern models suggest a peak about 1000 times brighter than a nova, the term ``kilonova'' seems apt \citep{2010MNRAS.406.2650M}.

For GRB\,050509B, \citet{2005ApJ...630L.117H}
used the model of
\citet{1998ApJ...507L..59L} and the upper limits shown in
Fig.~\ref{f:lightcurves} to constrain the fraction $f$ of the rest energy that
goes into radioactive decay (assuming $z=0.225$)
\citep[see also][]{2010MNRAS.404..963K}.
For a kinetic energy of order $10^{51}$ erg,
the approximate upper limit was found to be $f = 10^{-5}$.
The most efficient conversion of nuclear energy to the
observable luminosity is provided by the elements with a decay timescale
comparable to the time it takes the expanding ejecta to become optically thin. In reality, there is likely to be a large number of nuclides with
a very broad range of decay timescales. Taking both into account, the \citet{2005ApJ...630L.117H} limit constrains the abundances
and the lifetimes of the radioactive nuclides that form in the
rapid decompression of nuclear-density matter -- they should be either
very short or very long so that radioactivity is inefficient in generating
a large luminosity. In other words, unless the intrinsic energy in the
outflow from GRB\,050509B were $\ll 10^{51}\;$erg,
most of this energy was in sub-relativistic
ejecta with a very small radioactive component during the optically thick
expansion phase.


\subsection{Long-duration GRBs}

The absence of a SN signature in their light curves has been proposed as a defining 
characteristic of short GRBs. However, the discovery of another class of 
SN-less GRBs, namely long-duration GRBs like GRBs\,060505 
\citep{2006Natur.444.1047F}, 060614 \citep{2006Natur.444.1047F,2006Natur.444.1050D,2006Natur.444.1053G}, and possibly 051109B \citep{Perley06_GCN5387} 
\citep[and, in retrospect, XRF 040701;][see below]{2005ApJ...627..877S}
with no SNe observed to deep limits, poses a challenge to an otherwise clean classification scheme. While there is a possibility of a chance coincidence of a GRB with a foreground galaxy 
\citep{2006ApJ...651L..85C,2008MNRAS.391..935C}, thereby giving the false impression of deep non-detections of SN light, the probability that all such associations are spurious becomes vanishingly small. One proposed resolution was to posit the $> 100$ s long GRB\,060614 as belonging to a ``short'' GRB population \citep{2007ApJ...655L..25Z}. 
While we do not favor such a classification \citep{2008AIPC.1000...11B},
it highlights the importance of this new class of objects for our understanding
of the GRB-SN connection.

GRB\,060505 ($z=0.089$) was a GRB with a duration of 4--5 s (depending on bandpass) and a non-zero spectral lag (namely, the delay of arrival of low energy photons with respect to higher energy photons) of
0.36 s, inconsistent with the lags of short GRBs
\citep{2008ApJ...677L..85M}.
It occurred in a star-forming region of a galaxy and exhibited all the 
characteristics of a normal long-duration GRB
\citep{2006Natur.444.1047F,2008ApJ...676.1151T,2009ApJ...696..971X}\footnote{GRB\,060505 has been suggested to be a member of the short GRB class, belonging to
the long tail of the duration distribution \citep[][]{2007ApJ...662.1129O}. 
This is supported by the fact that it is an outlier of the
so-called Amati relation \citep{2007A&A...463..913A} just like short GRBs.
Note however that GRB\,980425 is also a significant outlier, as is GRB\,031203
if the {\it Integral} peak energy is adopted, c.f.\ Table \ref{t:five}.}.
Still, no SN was detected down to a limit 100 times fainter than 
SN\,1998bw \citep[][Fig.~\ref{f:lightcurves}]{2006Natur.444.1047F}.

This opens up the possibility that some long GRBs either do not produce SNe at all, or produce very faint SNe, or that no radioactive material is formed or ejected. 
It is possible that these constitute fall-back SNe that produce a black hole without 
forming an accretion disk in which the Ni can form 
\citep{1993ApJ...405..273W,2006ApJ...650.1028F,2010ApJ...719.1445M} or that 
the energy deposition is too low when the jet penetrates the star 
\citep{2006NCimB.121.1207N,2007ApJ...657L..77T}. Such events may finally  represent the incarnation of the original ``failed Ib'' SN model 
\citep{1993ApJ...405..273W}.

The true fraction of such SN-less GRBs is unknown. It is noteworthy that 
several XRFs have been reported as showing no SNe 
\citep{2005ApJ...627..877S,2005ApJ...622..977L}.
Unfortunately, most of these have no measured redshift and were not sampled at the 
expected peak of the SNe, so the limits are strongly model dependent. The 
most promising candidate is XRF\,040701, which occurred in a galaxy at $z=0.21$ 
\citep{2005ApJ...627..877S}.  {\it HST} observations around the time of the expected SN peak revealed no bump to 6 magnitudes fainter than the peak brightness of 1998bw. \citet{2005ApJ...627..877S} adopted the 
\citet{1995A&A...293..889P} relation to infer an extinction of $A_V < 2.8$ mag 
from the constraints on the equivalent column density of neutral hydrogen 
$N_H < 5\times 10^{21}$ cm$^{-2}$ as measured from the soft X-ray emission in the {\it Chandra} 
spectrum.  
However, a study of 28 {\it Swift} bursts with
joint observations of $A_V$ and $N_H$ \citep{2010MNRAS.401.2773S} suggests
an equivalent constraint of $A_V < 1$ mag, 
indicating that the limit on the absence of an SN related to XRF\,040701 is indeed very strong,
similar to those of GRB\,060505 and GRB\,060614. 
In this connection we note that the soft X-ray
emission following some short GRBs could mimic an XRF if the initial
short spike is too hard to be detected by {\it Swift}/BAT
\citep{2005Natur.437..859H}. 

\section{Other properties of GRB SNe}

Below we briefly address other important aspects of the GRB-SN connection.
The basic properties of the GRBs, their SNe, and their host galaxies
are summarized in Table~\ref{t:five}.

\subsection{Host galaxies}

\begin{figure}
    \centering
    \includegraphics[scale=0.55]{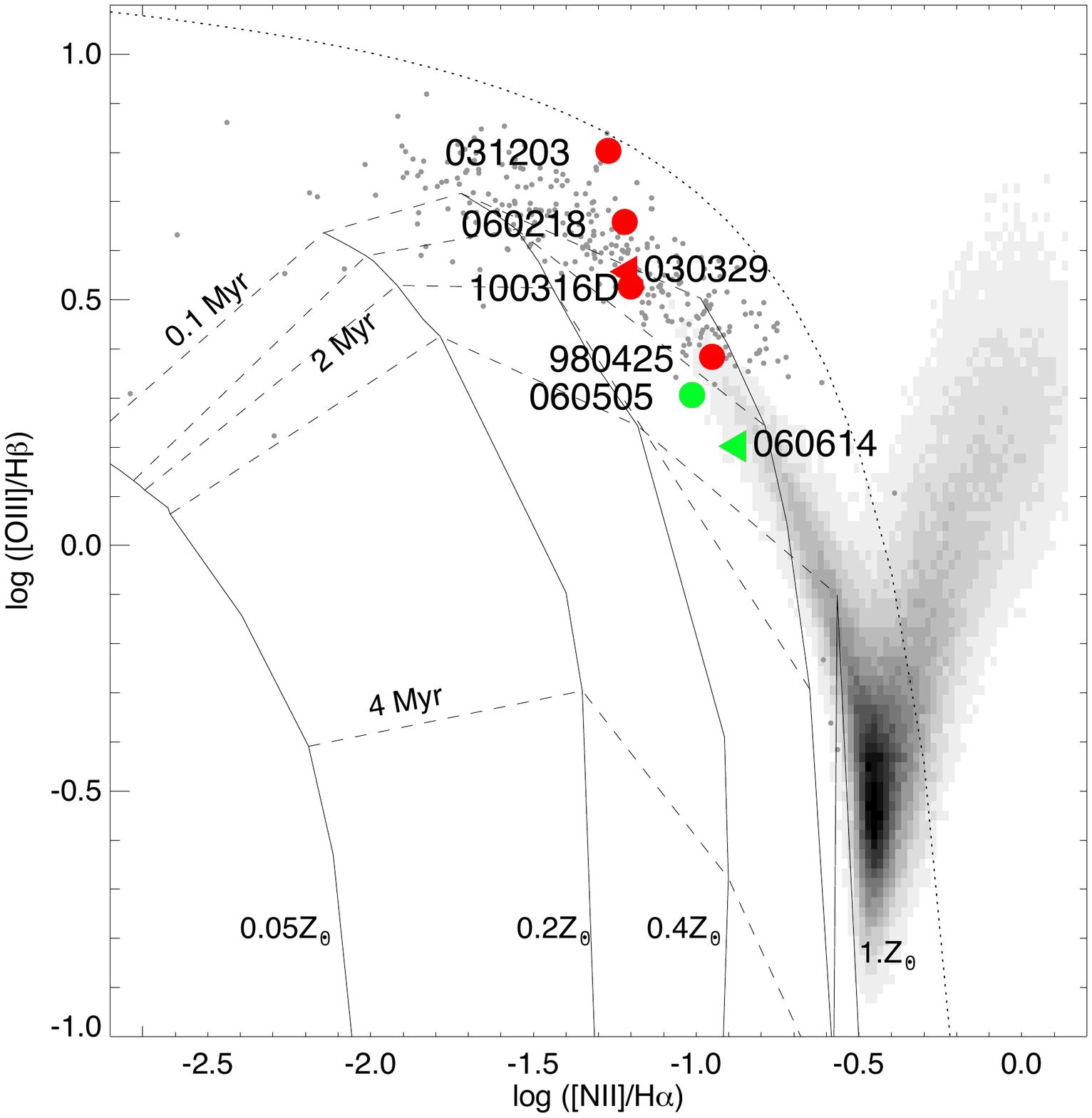}
    \caption{Host galaxies of GRB-SNe have sub-Solar metallicities.
The diagram \citep[adapted from][]{christensen08} shows the emission-line
ratios [O {\sc iii}]/H$\beta$ vs.\ [N {\sc ii}]/H$\alpha$ for GRB-SN sites
(full symbols) and SN-less GRBs (open symbols, cf.~Fig.~\ref{f:lightcurves}).
The line ratios for GRB\,100316D are for knot A in the host galaxy 
\citep{2011MNRAS.411.2792S}. Evolutionary 
models linking emission-line ratios at ages from 0.1 to 5 Myr for 
0.05, 0.2, 0.4 and Solar metallicities 
\citep[$Z_\odot = 0.016$, ][]{2005ASPC..336...25A}
are plotted as solid lines. The dotted curve denotes the 
separation of pure star-forming galaxies from AGN-dominated emission-line 
ratios. The greyscale area represents SDSS galaxies and the small grey dots 
represent SDSS galaxies with metallicities between 0.03 and 0.7 Solar. 
Two hosts with upper limits for [N {\sc ii}] are represented by triangles.
    }
    \label{f:hosts}
   \end{figure}

The host galaxies of GRB-SNe generally share the properties found
for GRB hosts (see Chapter 13), namely they are sub-luminous, blue,
young, and star-forming. In particular, the host galaxy of GRB\,980425
(Fig.~\ref{f:1998bw_discovery}) has been the subject of fairly intense 
scrutiny because of its proximity
\citep{2000ApJ...542L..89F,2006A&A...447..891F,2006A&A...454..103H,christensen08,michalowski09}.
It has been suggested that the host galaxies of GRBs
 \citep{2003A&A...406L..63F,2006AcA....56..333S,2006Natur.441..463F}
and GRB-SNe
\citep{2008AJ....135.1136M}
occur in low-metallicity environments. Fig.~\ref{f:hosts} shows
the location of GRB-SN hosts in a line-ratio diagram
\citep{christensen08}. As noted by \citet{2008AJ....135.1136M}, GRB-SNe occur in 
environments that are of systematically lower metallicity than broad-lined
                 SN Ic with no known association to GRBs.
The host galaxies of spectroscopic GRB-SNe discovered so far clearly have sub-Solar, but not very low, metallicities \citep{2011arXiv1101.4418L}. 

\subsection{Timing}

\citet{1998Natur.395..672I} found from modelling the light curve and spectra of
SN\,1998bw that the time of core collapse could be set to coincide 
with the detection of GRB\,980425 to within +0.7/$-2$ days; a similar conclusion was reached by \citet{1998Natur.395..663K}.
\citet{2003Natur.423..847H}
attempted to date SN\,2003dh spectroscopically, under the assumption of 
its parallel spectral evolution with SN\,1998bw and
concluded that SN\,2003dh began $\pm$2 days relative to GRB\,030329
(to this should be added the above uncertainty of +0.7/$-2$ days of the 
SN\,1998bw dating). Finally, the near-simultaneous detection of GRB\,060218
and what was interpreted as the shock breakout of SN\,2006aj firmly
linked the times of explosions of this GRB and its associated SN
\citep{2006Natur.442.1008C}. Supranova models have simply not withstood the test of observations with new events.

\subsection{Asymmetry and asphericity}
Asphericity may be a common feature of core-collapse SNe 
\citep[e.g.,][]{2006Natur.440..505L,2008Sci...319.1220M}
and hence GRB-SNe. There is some observational evidence that GRB-SNe
may be aspherical from the shapes and relative expansion velocities of 
emission lines, including highly peaked oxygen lines in nebular spectra 
\citep[e.g.,][]{2001ApJ...559.1047M,2006ApJ...645.1331M,2007ApJ...661..892M,2007RMxAC..30...23M}. Similarly, 
imaging polarimetry and spectropolarimetry point to aspherical emission
\citep[e.g.,][]{2003Natur.426..157G,2006A&A...459L..33G,2007A&A...475L...1M}, 
but the exact interpretation of these results and their implications for
the shape and internal composition of GRB-SNe is still under debate
\citep[e.g.,][]{2008ApJ...687L...9M,2007A&A...475L...1M,2008Sci...319.1220M}.

\subsection{Circumburst environment}

A massive-star progenitor should expel mass pre-explosion and a signature of a ``wind-stratified'' medium should, in principle, be manifest in the observations of the afterglow.
Despite this nominal expectation, few GRB afterglows  are significantly better fit by a medium consistent with a constant mass loss (circumburst mass density $\rho \propto r^{-2}$, with $r$ as the distance to the center of the explosion; \citealt{cl99,pk02}) than a constant-density medium. There are some reasonably strong cases for a $\rho \propto r^{-2}$ circumburst medium \citep{pbr+02} but, in general, the nature of the density profile proxied by the afterglow is both model-dependent and highly coupled to measurements of other afterglow-specific parameters (e.g., \citealt{2003ApJ...597..459Y,2008ApJ...672..433S}). 
Given the preponderance of direct evidence linking GRBs and massive star deaths, we feel that the lack of a tell-tale wind signature must imply an incomplete understanding of afterglow emission mechanisms and/or the details of the progenitor evolution rather than serve as disconfirming evidence for the connection. Indeed,  this lack has been accommodated in models as due to the homogenizing influence of the wind reverse shock of the density surrounding the star \citep{2001grba.conf..306W,2005ApJ...631..435R} and varying mass-loss histories of the progenitor \citep{2006A&A...460..105V}.  

GRB\,021004 showed absorption systems at high-velocities relative to the host galaxy in the afterglow, suggestive of a Wolf-Rayet wind \citep{2003ApJ...595..935M,2003ApJ...588..387S,2010A&A...517A..61C}; note, however, that it has been argued by \citet{2007ApJ...663..420C} that the absorption systems are unlikely to have arisen from within the circumburst environment. In general,
the absence of high-velocity signatures in absorption spectra of several other events have been explained as due to the photo-destructive power of the early afterglow \citep{2006ApJ...648...95P,2007ApJ...663..420C}.  Interestingly, an initial study of high-ionization species (of, e.g., nitrogen), expected to survive the early afterglow, appear to favor low-velocity circumburst outflow pre-explosion \citep{2008ApJ...685..344P}.

Finally, in a study of the high signal-to-noise {\it Swift} X-ray spectrum
of GRB\,060218,
\citet{2008ApJ...683L...9C}
inferred CNO abundances much larger than the Solar value, presumably due 
to enriched circumburst material. Interpreting the relative abundance ratios 
of CNO in the context of the models of \citet{2006A&A...460..199Y}
indicates that the required mixing limits massive progenitor star models
to have sub-Solar metallicity and rapid rotation.

\def\Kaneko{1}
\def\Watson{2}
\def\Woosley{3}
\def\Mazzali{4}
\def\Ceron{5}
\def\Sollerman{6}
\def\Modjaz{7}
\def\Prochaska{8}
 \begin{table}
  \caption{Properties of the Associated GRBs and SNe.}
    \begin{tabular}{l|lllll}
     \hline \hline
GRB                                 &{\bf 980405}&{\bf 030329}&{\bf 031203}&{\bf 060218} & {\bf 100316D}\\
     \hline
SN designation                      &1998bw&2003dh&2003lw&2006aj&2010bh        \\
$z$                                 &0.0085&0.1685&0.1055&0.0334&0.0591       \\
     \hline
\multicolumn{5}{|c|}{High-Energy and Afterglow Properties} \\
\hline
$T_{90}$ (s)                        &$34.9\pm3.8$  & $22.9$ &  $37.0\pm1.3$& $2100\pm100$ & 1300        \\
$S_X/S_\gamma$                      &0.58  & 0.56 & 0.49(4$\pm$2)& 3.5 & 1.56 \\
$E_{\rm peak}$ (keV)                &$122\pm17$   & $70\pm2$   &$>71$($<20$)& $4.7\pm1.2$ & $18^{+3}_{-2}$   \\
$E_{\gamma,\rm iso}$ ($10^{51}$ erg)&9$\times10^{-4}$& 13   &  0.17& 0.04 & $0.06$    \\
$E_{\gamma}$ ($10^{51}$ erg)        &$<9\times10^{-4}$&0.07--0.46&$<0.17$&  $<0.04$ &0.0037--0.06  \\
     \hline
	\multicolumn{5}{|c|}{Supernova Properties}\\
\hline
$M_{\rm bol,peak}$ (mag)            &$-18.6$&$-18.7$&$-18.9$&$-18.2$&$-17.5$\\
$v_{\rm exp}$ at 10 d ($10^3$ km s$^{-1}$) & 24   & 29   &  21  & 19         \\
M($^{56}$Ni) (M$_\odot$)            &0.38--0.48&0.25--0.45 &0.45--0.65  & 0.20--0.25 & $\sim 0.10$        \\
M$_{\rm ejecta}$ (M$_\odot$)            & 10$\pm$1    & 8$\pm$2  &  13$\pm$2 &  2$\pm$0.5 & $\sim 3$        \\
M$_{\rm ZAMS}$ (M$_\odot$)          &35--45     &25--40   &40--50    & 20--25          \\
$E_{\rm SN}$ ($10^{51}$ erg)               & 50$\pm$5   & 40$\pm$10 & 60$\pm$10  &  2$\pm$0.5 & $\sim 10$         \\
     \hline
		\multicolumn{5}{|c|}{Host Galaxy Properties} \\
	\hline
$M_{\rm B,host}$ (mag)              &$-17.7$&$-16.5$&$-21.0$ &$-15.9$&$-18.8$  \\
H$\alpha$ SFR (M$_\odot$ yr$^{-1}$) &0.23   & 0.6& 12.3  & 0.065  & $>0.17$  \\
M$_*$ ($10^9$ M$_\odot$)            &1.1    & 1.5 & $\sim$1   & 0.05      \\
SSFR (Gyr$^{-1}$) &0.21   & 0.4& 1  & 1.3    \\
Metallicity & 8.25--8.39 & 7.8 & 8.12 & 8.0 &8.23 \\
     \hline \hline
    \end{tabular}
  \label{t:five}
High-energy and afterglow properties: taken from
\citet{2007ApJ...654..385K} and references therein,
supplemented by 
\citet{2004ApJ...605L.101W,2006ApJ...636..967W,2011MNRAS.411.2792S},
and 
\citet{2006ARA&A..44..507W} and references therein.
The values in parentheses for $S_X/S_\gamma$ and $E_{\rm peak}$
given for 031203 are inferred by
\citet{2004ApJ...605L.101W} using the X-ray dust echo.
Supernova properties: taken from
\citet{2007RMxAC..30...23M} and references therein.
Host galaxy properties: absolute magnitudes from
\citet{2005NewA...11..103S}, \citet{2007A&A...464..529W}, and
\citet{2006A&A...454..503S},
stellar masses from
\citet{2010ApJ...721.1919C}, and
star-formation rates from
\citet{christensen08},
\citet{2005A&A...444..711G},
\citet{2004ApJ...611..200P}, and
\citet{2007A&A...474..815M}. Metallicities here are 12 + log(O/H), as compiled in \citet{2007A&A...474..815M}.
Values for GRB 100316D are from \citet{2011MNRAS.411.2792S} and
\citet{2011AN....332..262B};
the metallicity in this case 
refers to knot ``A'' close to the SN site. 
Key: 
``$T_{90}$'': Duration of 90\% of the event in $\gamma$ rays.
``$S_X/S_\gamma$'': Ratio of fluence in 2--30 keV band to fluence in 30--400 keV band. A ratio larger than 1 indicates an XRF.
``$E_{\rm peak}$'': Photon energy where $\nu f_\nu$ peaks.
``$E_{\gamma,\rm iso}$'': Isotropic equivalent energy output in $\gamma$ rays (i.e., not corrected for beaming).
``$E_{\gamma}$'': Energy output in $\gamma$ rays (i.e., corrected for beaming).
An upper limit indicates that no beaming was measured.
``$v_{\rm exp}$'': Photospheric expansion velocity.
``M($^{56}$Ni)'': Mass of $^{56}$Ni synthesized.
``M$_{\rm ejecta}$'': Mass of the ejecta.
``M$_{\rm ZAMS}$'': Mass of the progenitor star (on the zero-age main sequence).
``$E_{\rm SN}$'': Energy associated with the outflow of the SN.
``SFR'': Star-formation rate.
``M$_*$'': Stellar mass.
``SSFR'': Specific star formation rate (SFR/M$_*$).
 \end{table}

\section{Outlook after a decade of the GRB-SN connection}

It now appears unassailably established that many (if not most) long-duration GRBs are connected with the death of massive stars: the same event that produces a GRB also produces a substantial mass in $^{56}$Ni and a large coupling of kinetic energy ($>10^{51}$ erg) to non-relativistic ejecta.  The pre-explosion jettisoning of hydrogen and helium appears to be an observational requirement of the progenitor (see Chapter 10 for a theoretical review). Still, we do not understand what sets a stripped-envelope GRB/SN progenitor apart from other SN progenitors which do not produce GRBs. 

From demographic studies of the relative rates of GRBs and SNe
(e.g.,
\citealt{1998ApJ...506L.105B,2003ApJ...599..408B,2004ApJ...607L..17P,2004ApJ...607L..13S,2006AIPC..836..380S})
it is clear that only a tiny fraction ($\sim$1\%) of core-collapse
events that produce Ibc SNe also produce a detectable GRB. Even Ic-BL
SNe, which themselves comprise a small fraction of the core-collapse
SN rate \citep{2007ApJ...657L..73G}, are seldom associated with a
GRB. 
 Since GRB jets are thought to be highly collimated (and
 relavistically Doppler beamed), whereas SNe are roughly isotropic, the
 difference in rates might simply be due to geometry and
 viewing-angle effects. \citet{2006ApJ...638..930S} found no evidence
 for highly energetic and off-axis components in  radio observations of
 Ic-BL SNe, statistically disfavoring the notion that all Ic-BL SNe
 also harbor GRB-like (``central engine driven'') features.  The
 recent inferences of mildly relativistic ejecta in the Type Ic SN\,2009bb
 \citep{2010Natur.463..513S} may suggest a
 central-engine driven origin. At minimum, it seems such objects
 support the notion of a continuum of energy and mass coupled to
 relativistic speeds across SNe.

History is some guide here: though the GRB associated with SN\,1998bw appeared to belong to a distinct subclass in terms of energetics, more recent events point to a continuum of energy coupled to the relativistic ejecta. Likewise, though it now appears that GRB\,060614 and GRB\,060505 represent a distinct subclass in terms of $^{56}$Ni production ($M_{V, {\rm peak}}>-13$\, mag), transition GRB-SNe with faint SNe ($M_{V, {\rm peak}}$ $\sim$ $-$16\, mag, c.f.\ Fig.~\ref{f:lightcurves}) may be found, pointing to a continuum of explosive nucleosynthesis. 
Indeed, there is evidence for very faint ($M_{V, {\rm peak}}$ $\sim$ $-$14) 
extragalactic events residing on the tail end of the $^{56}$Ni production 
distribution in core-collapsed Type II SNe 
\citep[][Fig.~1]{2007Natur.449....1P}, although the existence of ultra-faint Type Ibc 
SNe is currently debated \citep{2009Natur.459..674V,2009AJ....138..376F,2010ApJ...719.1445M}. This
supports the physical intuition that massive stars which explode can produce 
a wide range of $^{56}$Ni mass with a broad range of explosion energies (c.f., \citealt{2007ApJ...661..892M}).

The distribution of available energy into  various channels (neutrinos, gravitational waves, $\gamma$ rays, particles, kinetic motion of the relativistic outflow, kinetic motion of the non-relativistic SN outflow) is far from understood.  There is reasonable evidence, from the metrics proxying the bulk kinetic energy of the afterglow ($E_K$) and the measures of energy promptly released in $\gamma$ rays ($E_\gamma$), to suggest an overall envelope in the energy budget of the explosions \citep{2003Natur.426..154B}. More recent work including very luminous {\it Fermi} GRBs \citep{2010arXiv1004.2900C}, however, suggests that $E_\gamma$ and $E_K$ are correlated and so the total budget may not be similar among all long-duration events.

In all cases where the energy associated with the non-relativistic outflow of the SN ($E_{\rm SN}$) has been modelled, it appears to be larger than $E_\gamma$ and $E_K$. To gain some insight into the energy partitioning, it is tempting to try to connect the amount of SN energy (and the other basic parameters $M_{\rm ejecta}$ and $^{56}$Ni mass) with the other measured energies associated with the GRB and afterglow. The danger is that observational biases that could induce apparent correlations are severe. For example, lower $E_\gamma$ events are more readily detected at low redshift, improving the SN detection possibility. Likewise, lower $E_K$ manifests itself as fainter afterglow, also improving SN detection odds. Indeed, a careful statistical study of the relationship between GRB-SN peak brightnesses and GRB energetics found no strong correlation \citep{2009MNRAS.393.1370K}. Still, this accounting for the global energy budget should continue to be pursued, especially in the next decade as gravitational-wave and neutrino detectors  begin to yield insight into the energy budgets of generic core-collapsed SNe in the local Universe.

The current spectroscopic evidence for GRB-SNe is not of high enough quality and the number of 
sample members is insufficient to address the issue of whether there are 
systematic differences between low-redshift (low $E_\gamma$) SNe and 
higher-redshift (high $E_\gamma$) SNe 
\citep{2005ApJ...627..877S,2000ApJ...537L.127S}.
Clearly, the group of 5 low-redshift GRBs ($z< 0.1685$) with spectroscopically secure SN identifications appear subenergetic in terms of beaming-corrected $E_\gamma$ than the majority of the 
higher redshift events. 
Establishing the
GRB-SN connection for higher-redshift, more energetic GRBs will bean observational challenge, requiring 30--40-m class telescopes. Other aspects of such an endeavor will be to determine the fraction of SN-less GRBs and their relation to the GRB properties and progenitors, as well as possibly finding the intermediate $^{56}$Ni production GRB-SN (which we might call ``gap'' GRB-SNe) discussed above 
(if indeed they exist). If
 some unidentified TeV sources in the Galaxy are associated with GRB remnants 
 \citep{2010ApJ...709.1337I}, then detailed chemical studies of the explosion 
 byproducts may provide new insight into the progenitors 
 (akin to late time studies of
  SN remnants; \citealt{2009AIPC.1111..307B}).

A hint that surprises may be in store are the very significant light curve
bumps in some GRB afterglows that do not fit our current understanding
of the GRB-SN connection:
\begin{itemize}
\item
GRB\,020305 showed a clear bump, both in the light curve and in the spectral
shape  \citep{2005A&A...437..411G}. Each of these were found to be consistent 
with (different) SN templates, but no single SN template could fit both 
features simultaneously.
\item
XRF\,030723 had a very significant and well-sampled bump showing a
clear reddening at the peak of the bump \citep{2004ApJ...609..962F}
and was well fitted by a SN model \citep{2004ApJ...612L.105T}. 
However, a reported possible simultaneous X-ray
increase \citep{2005ApJ...621..884B} shed doubts on the SN origin of the
light curve bump.
\end{itemize}

What are these bumps? Are they indications of SNe with different properties
from those discussed so far or do they point to mechanisms for producing
SN-like bumps but of an entirely different origin
\citep[e.g., dust echoes,][]{2000ApJ...534L.151E}? We note that in all cases, more limited observational information  (e.g., only one band, no X rays) ironically could have led to a higher confidence grade in Table 9.1. And it highlights the importance of
spectroscopic confirmation of SNe, as emphasized in this chapter.

As unconventional SNe may be related to normal GRBs, unconventional
high-energy events may be related to normal SNe. The discovery of the Type Ib
SN 2008D related to an X-ray transient (i.e., not a GRB)
\citep{2008Natur.453..469S,2008Sci...321.1185M,2009ApJ...692L..84M,2009ApJ...702..226M}
suggests that SNe may reflect
a range of high-energy properties, not only GRBs. It opens up
the possibility that all SNe may be related to detectable 
high-energy events and that GRBs may be only one possible manifestation
of these, perhaps as part of a range of high-energy properties
\citep{2008arXiv0801.4325X,2008MNRAS.388..603L}.

The discovery of SN-less GRBs suggests that massive stars may die
without sending high-energy electromagnetic signatures of their deaths, e.g., 
off-axis events, and with
no bright SN event. The fraction of such events is unknown. Even in the most optimistic cases \citep{2002ApJ...565..430F,2006PhRvL..96t1102O}, gravitational-wave and neutrino signatures would only be detectable for the very nearest (and, hence, uncommon) events. However, there may be a case for mapping and subsequently revisiting/monitoring all the massive stars in the
nearby Universe with {\it HST}, {\it JWST}, or {\it Euclid}. 
This would give an independent handle on the
fraction of disappearing stars \citep{2008ApJ...684.1336K}, at least in the types of galaxies
surveyed. 
In a complementary approach, the search for kilonovae or
similar new types of radioactive-powered events would be important to map all the
possible cosmic explosions, including those resulting from (short) GRBs.

The connection between some GRBs and massive stellar deaths holds great importance in the utility of GRBs as probes of the distant Universe. First, massive stars trace the location of stellar nurseries at high redshift.  Though the first few seconds of afterglow light are generally thought capable of destroying most spectroscopic signatures of the circumburst and molecular cloud environments, spectroscopy of afterglow in some rare cases should inform the nature of these close-in environments \citep{2008ApJ...685..344P,2009ApJ...691L..27P}.  Second, the connection, especially emerging in low-metallicity hosts, implies that GRBs are tracers of instantaneous star formation (e.g., \citealt{2010ApJ...711..495B}). Not only can high-resolution spectroscopy directly constrain individual star-formation histories of hosts \citep{2009ApJ...693.1236C}, the comparative demographics of detections at various wavebands (particularly, $\gamma$ rays, X rays and optical/IR) should supply direct constraints on the fraction of obscured star formation in the Universe (e.g., \citealt{dfk+03}). This is one of the great hopes for the importance of the so-called ``dark bursts'' (\citealt{2009AJ....138.1690P}; see Chapter 6). Last, and perhaps most compelling, is that since GRBs appear to arise from low-metallicity, high-mass stars, they should exist beyond the redshifts of furthest known objects today; indeed this notion stands on rather firm observational ground now with the identification of several events above redshift $z=6$. GRBs, then, are likely the most recognizable events in the very early Universe, detectable by the next-generation of satellites and signposts (for {\it JWST} and the next generation 30-m telescopes) to the first galaxies. Even moderate-resolution spectroscopy of GRB afterglows beyond $z=7$ (e.g.\ with GRB\, 090423 at $z=8.2$; \citealt{2009Natur.461.1254T}) would be unique {\it in situ} tracers of the cosmo-chemical evolution of the infant Universe (e.g., \citealt{2009astro2010S.199M}).

\section*{Acknowledgements}

We thank Elena Pian and Milena Bufano for their input to
Fig.~9.2 and Table~9.2, and Lise Christensen for producing 
Fig.~9.5. We thank
Bethany Cobb, 
Massimo Della Valle,
Johan Fynbo,
Javier Gorosabel,
Justyn Maund,
Maryam Modjaz, 
Adam Morgan, 
Daniel Perley,
Jesper Sollerman,
Rhaana Starling,
and Dong Xu
for a close reading of a draft of this chapter.
The Dark Cosmology Centre is funded by the Danish National Research
Foundation. J. S. B. was partially supported by a grant NSF/AST-1009991.

\bibliography{journals_apj,cup}

\end{document}